\documentclass[aps,prl,reprint,superscriptaddress]{revtex4-2}
\usepackage{hyperref}
\hypersetup{
    colorlinks = true,
    linkcolor = [rgb]{0.70, 0.13, 0.13},
    citecolor = [rgb]{0.13, 0.55, 0.13},
    urlcolor = [rgb]{0.25, 0.41, 0.88}
}
\usepackage{amsmath}
\usepackage{amsfonts}
\usepackage{amssymb}
\usepackage{pifont}
\usepackage{mathtools}
\usepackage{bm}
\usepackage{cases}
\usepackage{braket}
\usepackage[version=4]{mhchem}
\usepackage{graphicx}
\begin{document}

\title{Fractionalization on the Surface: Is Type-II Terminated $1T$-TaS$_2$ \\ Surface an Anomalously Realized Spin Liquid?}
\author{Chao-Kai Li}
\thanks{These authors contributed equally.}
\affiliation{Department of Physics and HKU-UCAS Joint Institute for Theoretical and 
Computational Physics at Hong Kong, The University of Hong Kong, Hong Kong, China}
\affiliation{The University of Hong Kong Shenzhen Institute of Research and Innovation, Shenzhen 518057, China}
\author{Xu-Ping Yao}
\thanks{These authors contributed equally.}
\affiliation{Department of Physics and HKU-UCAS Joint Institute for Theoretical and 
Computational Physics at Hong Kong, The University of Hong Kong, Hong Kong, China}
\affiliation{The University of Hong Kong Shenzhen Institute of Research and Innovation, Shenzhen 518057, China}
\author{Jianpeng Liu}
\affiliation{School of Physical Science and Technology, ShanghaiTech University, Shanghai 200031, China}
\affiliation{ShanghaiTech Laboratory for Topological Physics, ShanghaiTech University, Shanghai 200031, China}
\author{Gang Chen}
\email{gangchen@hku.hk}
\affiliation{Department of Physics and HKU-UCAS Joint Institute for Theoretical and 
Computational Physics at Hong Kong, The University of Hong Kong, Hong Kong, China}
\affiliation{The University of Hong Kong Shenzhen Institute of Research and Innovation, Shenzhen 518057, China}

\begin{abstract}
The type-II terminated $1T$-\ce{TaS2} surface of a three-dimensional $1T$-\ce{TaS2} 
bulk material realizes the effective spin-$1/2$ degree of freedom on each 
David star cluster with ${\mathcal{T}^2=-1}$ such that the time-reversal symmetry 
is realized anomalously, despite the fact that bulk three-dimensional $1T$-\ce{TaS2} material 
has an even number of electrons per unit cell with ${\mathcal{T}^2=+1}$. 
This surface is effectively viewed as a spin-$1/2$ triangular lattice magnet, 
except with a fully gapped topological bulk. 
We further propose this surface termination realizes a spinon Fermi surface 
spin liquid with the surface fractionalization but with a nonexotic 
three-dimensional bulk. We analyze possible experimental consequences, 
especially the surface spectroscopic measurements, 
of the type-II terminated surface spin liquid. 
\end{abstract}

\maketitle

There has been great interest in the theoretical community to 
classify distinct topological phases with or without 
symmetry~\cite{PhysRevB.80.155131,PhysRevD.88.045013,PhysRevB.87.155114,PhysRevB.84.235141,PhysRevX.3.011016,PhysRevB.83.035107,
PhysRevX.6.011034,PhysRevB.87.104406,PhysRevB.91.134404,PhysRevB.87.155115}. 
These include classifying different symmetric spin liquids and
classifying symmetry-enriched topological orders and symmetry 
protected topological phases. Questions like where to realize the 
classified topological states were raised and partially 
understood~\cite{PhysRevB.87.104406}. 
It was understood that two-dimensional (2D) 
spin liquids (and/or 2D intrinsic topological orders) 
with certain symmetry fractionalizations cannot be 
realized in strictly 2D systems~\cite{PhysRevB.87.104406,
PhysRevB.91.014405}. 
Instead, they may be realized on the 2D surface of the 
three-dimensional (3D) symmetry protected topological states 
where the symmetry is realized anomalously~\cite{PhysRevB.87.104406}. 
Likewise, similar results were established in the study of time-reversal 
symmetric 3D U(1) spin liquids where time-reversal symmetry is 
realized anomalously for the spinons and supports 
gapless surface liquid states~\cite{PhysRevX.6.011034,Pesin2010}. 
Despite the interesting theoretical advances, such anomalous realizations of the low-dimensional spin liquids on the surface of the high-dimensional topological states, including those that can in principle be realized in strictly low-dimensional systems, 
have not yet been achieved.

\begin{figure}[b]
	\includegraphics[width=8.6cm]{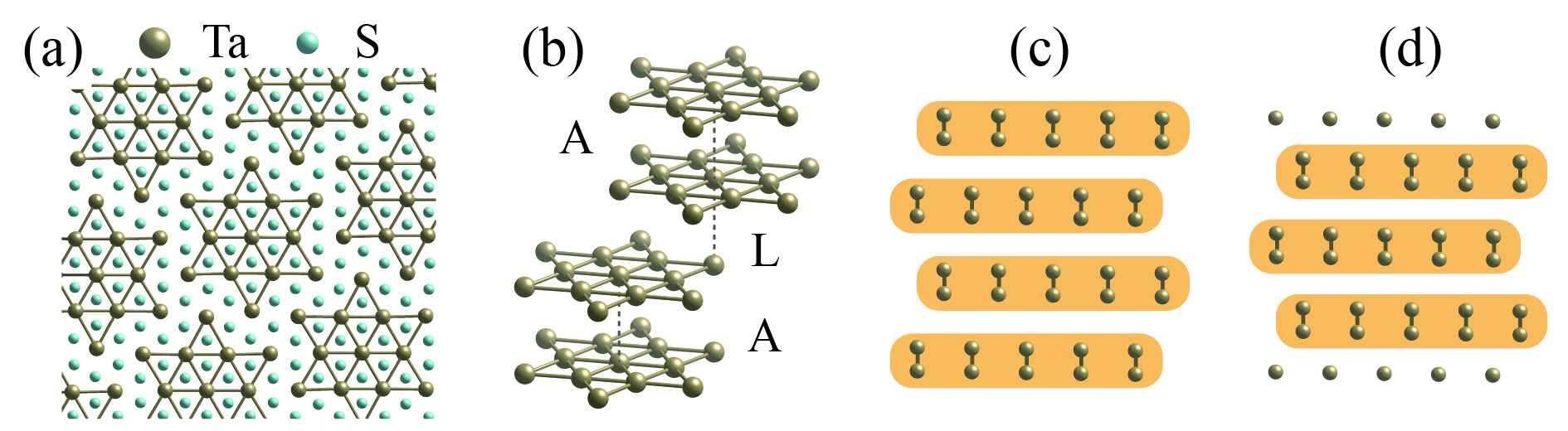}
	\caption{(a) The Star of David clusters. 
	After the charge-density-wave transition, the Ta--Ta bonds are contracted, forming a ${ \sqrt{13} \times \sqrt{13} }$ supercell structure. 
	(b) The out-of-plane stacking structure. 
	The $A$-type and the $L$-type stacking occur alternately, which is referred to as the $AL$ stacking. 
	(c) Schematic of the type-I termination. 
	(d) Schematic of the type-II termination. }%
	\label{fig:lattice_structure}
\end{figure}

In this Letter, we turn to the material's side and point out that the 
3D multilayer-stacked $1T$-\ce{TaS2} can be a candidate to anomalously 
realize the 2D spin liquid on its surface.   
This system has several interesting properties that differ from 
the conventional magnets. 
With the 3D multilayer stacking, the system develops the  
dimerization between neighboring layers along the $z$ direction 
such that there exist even number of layers in each unit cell 
(see Fig.~\ref{fig:lattice_structure}). 
The system develops a ${\sqrt{13} \times \sqrt{13}}$ 
charge-density-wave order at low temperatures, 
and the \ce{TaS2} layer distorts to form cluster units with 
the shape of the Star of David. There exist $13$ electrons in 
each Star of David, and thus one unpaired electron in the cluster unit.
As the 3D system has even \ce{TaS2} layers in the unit cell, 
the whole system is connected to a band insulator with a band gap, 
and we do not expect exotic quantum phases to emerge in the bulk. 
This is actually compatible with the Lieb-Schultz-Mattis-Oshikawa-Hastings (LSMOH) theorem that states the possibility of topological order and spin liquids 
in insulators with odd electron fillings~\cite{PhysRevLett.84.3370,PhysRevB.69.104431,LIEB1961407}. 
The surface of this material, however, makes a difference. 
Because of the dimerization, there exist two distinct surface 
terminations~\cite{Butler2020} (see Fig.~\ref{fig:lattice_structure}). 
The type-I (type-II) surface terminates at the end (middle) of the dimer.
In this Letter, we provide a theoretical understanding of two different surfaces 
and argue that the type-II surface is an anomalously realized spin liquid.

The observation is to consider the 1D chain built from the coupling 
of the Star of David cluster along the $z$ direction. 
Once one turns on the electron hopping along the chain, 
the model becomes a spinful version of the Su-Schrieffer-Heeger (SSH) 
model~\cite{RevModPhys.60.781} if one neglects the displacement of 
the dimers from the neighboring layers. 
As shown in Figs.~\ref{fig:lattice_structure}(c) and~~\ref{fig:lattice_structure}(d), the van der Waals bonds between the neighboring layers 
in the same dimer (in orange) are relatively strong, while those between different 
dimers are weak. 
The 1D chain supports gapless end states if the system terminates 
at the middle of the strong bond just like the type-II surface termination.  
Because of the spin part, the end states form a Kramers doublet with $\mathcal{T}^2=-1$. 
As each David star hosts one unpaired electron, in the real space picture, two electrons from the strong bond pair up, leaving the electron on the end of the chain unpaired. 
When the 3D system is formed as in Fig.~\ref{fig:lattice_structure}(d), the bulk is a band insulator while the surface is a 2D metal with an electron Fermi surface in our band structure calculation without including the strong correlation effect. 
When the electron correlation is included, the charge of the surface electron is clusterly localized in the David star, and the surface becomes Mott insulating. 
The remaining spin-$1/2$ degrees of freedom form a spin liquid. 
This surface spin-$1/2$ moment is anomalously realized, 
and so is the surface spin liquid. 
If the top surface of the system is type I and the bottom surface 
is type II, this ``thick'' 2D system would have an odd electron filling, 
and the presence of surface spin liquid is compatible with the LSMOH theorem.

%\emph{Surface states for type-II termination.}---

\begin{figure}
	\includegraphics[width=8.6cm]{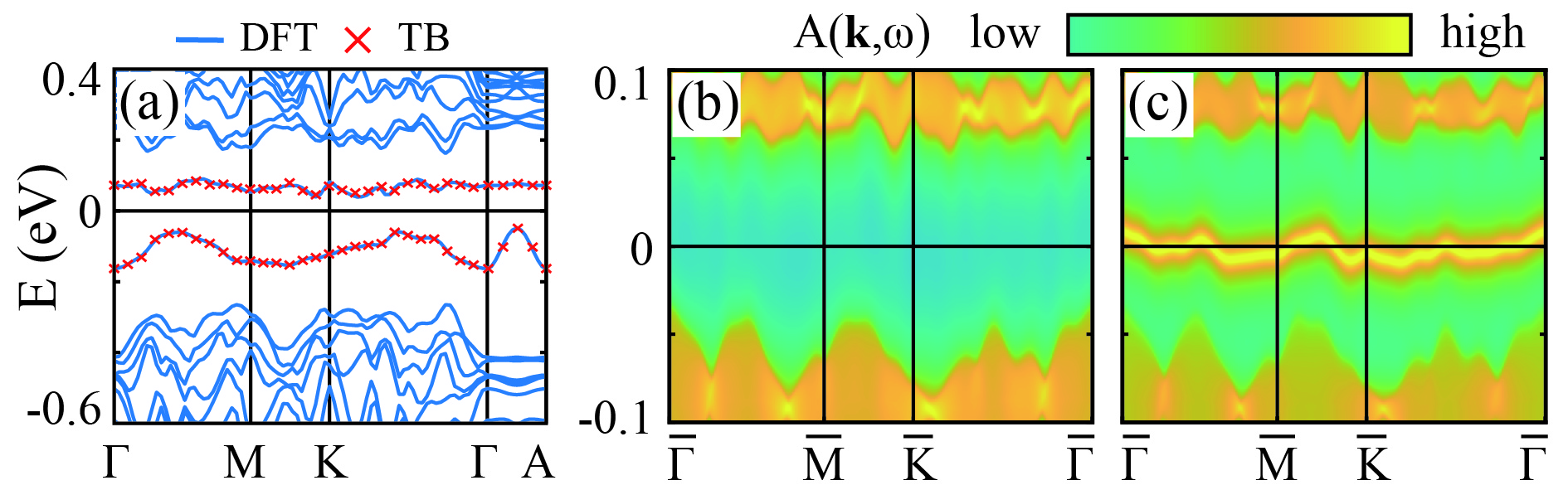}
	\caption{
	(a) The band structure of bulk $1T$-\ce{TaS2} along the high-symmetry lines in the Brillouin zone of the primitive cell. 
    The DFT and the tight-binding model results are shown by blue curves and red crosses, respectively. 
    The spectral functions $A({\bm{k}},\omega)$ of the type-I and type-II terminated semi-infinite models are shown in (b) and (c), respectively.}%
	\label{fig:DFT_results}
\end{figure}

The electronic structures of $1T$-\ce{TaS2} are calculated by a first-principles method within the Kohn-Sham scheme of density functional theory (DFT)~\cite{PhysRev.136.B864,PhysRev.140.A1133}, as implemented in the \textsc{Quantum} ESPRESSO package~\cite{Giannozzi2009,Giannozzi2017}. 
Details can be found in the Supplemental Material (SM)~\cite{SM}.\nocite{PhysRevB.78.045109,PhysRevLett.77.3865,PhysRevB.44.6883,PhysRevB.86.245102,PhysRevLett.118.087201,DalCorso2014,pslibrary,Sunko2020,Butler2020}
We use the $AL$ stacking structure [see Fig.~\ref{fig:lattice_structure}(b)], which was found to have the lowest energy previously~\cite{PhysRevB.98.195134,PhysRevLett.122.106404,Wang2022}. 
This stacking sequence leads to dimerization along the $z$ direction.
The resulting Ta--Ta bond lengths in the star varies from $3.17$ to $3.28\ \mathring{\text{A}} $. 
The interlayer distance is around $6.7 \ \mathring{\text{A}}$, which is larger than the experimental value $5.92 \ \mathring{\text{A}}$~\cite{Givens1977}. 
Adding the van der Waals force corrections can get a smaller interlayer distance, but 
does not qualitatively change the surface states shown below
due to the topological nature of the SSH model. 
Following the convention in the literature~\cite{PhysRevB.98.195134,PhysRevLett.122.106404}, the bulk band structure is plotted along the high-symmetry 
lines in the Brillouin zone of the 
${1 \times 1 \times 1}$ primitive cell~\cite{SM}, as shown in Fig.~\ref{fig:DFT_results}(a). 
As a result of the dimerization, 
the bulk band of the dimerized \ce{TaS2} is gapped. 
To obtain the low-energy model, we select the conduction and 
the valence band around the Fermi level, and construct the maximally localized 
Wannier functions using the \textsc{Wannier90} package~\cite{PhysRevB.56.12847,Pizzi2020}. 
The band structure of the resulting two-band tight-binding (TB) model is plotted by red crosses in Fig.~\ref{fig:DFT_results}(a). 
It can be seen that the TB band structure well reproduces the DFT result.

To investigate the surface states of different terminations, we calculate the 
Green's function of the semi-infinite model based on the bulk two-band TB model, 
utilizing an iterative procedure~\cite{Sancho1985} implemented in the 
\textsc{WannierTools} program~\cite{Wu2018}. 
The corresponding spectral functions for the type-I and type-II terminations 
are shown in Figs.~\ref{fig:DFT_results}(b) and~\ref{fig:DFT_results}(c), 
respectively. It is clearly seen that the type-I surface is gapped. 
In contrast, for the type-II termination, although the bulk is a band insulator, 
there is a surface band locating in the bulk gap. 
We have verified that the states on the surface band are indeed located 
on the surface (see the SM~\cite{SM}).
This surface band is half filled, making the type-II surface a two-dimensional metal 
if we neglect the correlation effects.

%\emph{Surface fractionalization and detection.}----

What would be the fate of this surface metal 
once the strong electron correlation is included? 
A large Kondo resonance was recently detected on the 
type-II surface upon the Pb doping~\cite{Shen2022}, 
indicating the presence of the surface local moments. 
Moreover, early works on the monolayer $1T$-\ce{TaS2} indicate a cluster Mott insulating state where the unpaired electron from each Star of David is Mott localized and forms the local spin-$1/2$ moment~\cite{Law6996,Klanjsek2017,PhysRevLett.121.046401,Manas-Valero2021,PhysRevLett.126.196406,Wang2020,Ruan2021,PhysRevLett.129.016402}. 
Based on these results, we propose, upon introducing 
the electron correlation, the surface electron is localized 
via Mott transition~\cite{SM} and forms the spin-$1/2$ Kramers 
doublet in each Star of David with ${\mathcal{T}^2=-1}$. 
This anomalously realized spin-$1/2$ moment on the surface rises 
from the combination of the type-II surface and the Mott localization. 
Effectively the surface becomes a spin-$1/2$ triangular lattice magnet, 
and this triangular lattice is the superlattice formed by the clusters. 
Like the situation for $1T$-\ce{TaS2} monolayer~\cite{Law6996,PhysRevLett.121.046401}, 
we expect that, the type-II surface realizes a spinon Fermi surface spin liquid.

This anomalously realized spin liquid, once formed on the type-II $1T$-\ce{TaS2} 
surface, could be detected indirectly by the angle-resolved photoemission. 
Here we propose an indirect spectroscopic detection scheme based on the 
surface doping or dosing of metallic layer. 
This differs from the direct spectroscopic measurement of the Mott insulating 
surface spin liquid~\cite{PhysRevB.87.045119}. 
We consider a conducting layer with itinerant electrons dosed 
on top of the type-II $1T$-\ce{TaS2} surface. 
A similar setup has been experimentally realized with graphene to investigate the proximity effect through scanning tunneling microscopy and spectroscopy quite recently~\cite{altvater2022revealing}. 
The conducting layer contains itinerant electrons that interact with the underlying unpaired electrons of the $1T$-\ce{TaS2} surface as 
\begin{align}\label{eq:electron}
    H_e = - & \frac{t_e}{2}\sum_{\langle{ij} \rangle \sigma} (c_{i\sigma}^\dagger c_{j\sigma} + \text{H.c.}) - \mu_e \sum_{i \sigma}c_{i\sigma}^{\dagger}c_{i\sigma}
    \notag  \\  
    + & \sum_{\langle{ik}\rangle \sigma \sigma'} K_{ik} (c_{i\sigma}^{\dagger}  \bm{\tau}_{\sigma \sigma'} c_{i\sigma'} ) \cdot \mathbf{S}_k,
\end{align}
where $c_{i\sigma}^{\dagger}$ ($c_{i\sigma}$) creates (annihilates) an electron with spin $\sigma$ at site $i$ on the conducting layer, $t_e$ is the hopping parameter, $\mu_e$ is the chemical potential and $\bm{\tau}$ are the Pauli matrices. 
$\mathbf{S}$ is the spin degree of freedom of the unpaired electrons on the cluster that is coupled to itinerant electrons via the Kondo-Hund coupling parameter $K$. 
Here the Kondo-Hund coupling is treated as a probing coupling for the emergent surface spinons. 
For the surface spin liquid of $1T$-\ce{TaS2}, the electron experiences a spin-charge separation where the charge sector is Mott localized while the spin sector is in the spin liquid phase. 
Without loss of generality, we consider a mean-field theory to incorporate the surface spinons that are described by a mean-field Hamiltonian 
\begin{equation}\label{eq:spinon}
    H_s = - \sum_{\langle {ij} \rangle \sigma} t_{s,ij} (f_{i\sigma}^{\dagger} f_{j\sigma} + \text{H.c.}) - \mu_s \sum_{i\sigma}f_{i\sigma}^{\dagger}f_{i\sigma},
\end{equation}
where $t_{s,ij}$ refers to the spinon hopping on the triangular lattice formed by the cluster, and $f_{i\sigma}^{\dagger}$ and $f_{i\sigma}$ (with ${\mathbf{S}_i = f_{i\sigma}^{\dagger} \frac{\bm{\sigma}_{\sigma\sigma'}}{2} f_{i\sigma'}}$) are
the spinon creation and annihilation operators on the cluster of 
the surface triangular superlattice. 
The chemical potential $\mu_s$ enforces the half filling constraint. 
For the uniform spinon hoppings with ${t_{s,ij}=t_s / 2}$, 
the mean-field theory describes the spinon Fermi surface 
U(1) spin liquid.

The coupling to the conduction electrons could significantly 
modify the effective removal operator of electrons in the 
Mott insulator and thus result in the energy gains of the photoelectric effect. 
Physically, the ejected electrons carry the information from both the conduction 
and the Mott insulating layers. 
In the weak Kondo-Hund coupling limit, it has been shown that, 
in addition to the pure electron part, there exists an intertwined spin and charge response at the leading order due to the convolution of the itinerant electron 
spectral function $A_e (\bm{k}, \omega)$ with the spin correlation function $\mathcal{S} (\bm{k}, \omega)$ of the Mott insulating layer~\cite{doi:10.1126/sciadv.aaz0611},
\begin{multline}
\label{eq:convolution}
    A(\bm{k},\omega\le0)\propto\int_{-\infty}^{0}\frac{d\omega'}{2\pi}\int\frac{d^{2}\bm{q}}{(2\pi)^{2}}K(\bm{q}) \\
    \times A_{e}(\bm{q},\omega')\mathcal{S}(\bm{k}-\bm{q},\omega-\omega'),
\end{multline}
where $ K(\bm{q}) $ describes the Kondo-Hund coupling between the itinerant electrons and the spins, and the spin correlation function $\mathcal{S}\left(\bm{q},\omega\right)=\int_{-\infty}^{\infty}d t e^{\imath \omega t}\left\langle \mathbf{S}_{-\bm{q}}\cdot\mathbf{S}_{\bm{q}}\left(t\right)\right\rangle$. Note that $ \mathcal{S}\left(\bm{q},\omega\right) $ vanishes for $ {\omega > 0} $~\cite{SM}. When the Mott insulating layer is a spin ordered state, the spin correlation $\mathcal{S} ( \bm{q}, \omega)$ diverges at the ordering wave vector $\mathbf{Q}$ and the zero frequency ${\omega = 0}$, resulting in $A(\bm{k},\omega<0) \propto K(\bm{k}-\mathbf{Q}) A_{e}(\bm{k}-\mathbf{Q},\omega)$. 
It can be immediately inferred that, the intertwined spectral function manifests as a replica of itinerant electron bands but is shifted by the wave vector 
$-\mathbf{Q}$ and modulated by the Kondo-Hund coupling in intensity.

\begin{figure}
    \centering
    \includegraphics[width=8.6cm]{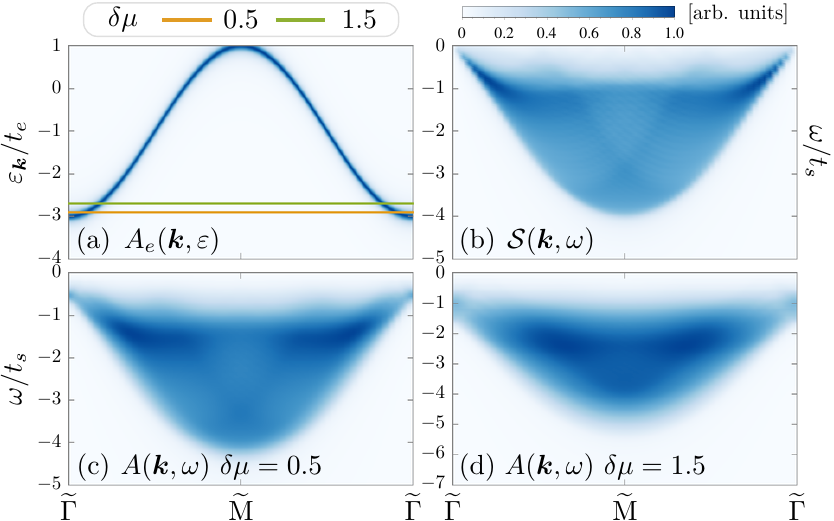}
    \caption{
    (a) The spectral function $A_e(\bm{k},\varepsilon)$ for conduction electrons. 
    Colored lines refer to Fermi levels for two dilute fillings where differences between the chemical potential $\mu_e$ and the band minimum $\varepsilon_{\text{min}}$ take $\delta \mu = 0.5$ and $1.5$, respectively. 
    (b) The spin correlation $\mathcal{S}(\bm{k},\omega)$ of U(1) spin liquid with the spinon Fermi surface. 
    (c),(d) The intertwined electron spectral function $A(\bm{k},\omega)$ for dilute fillings shown in (b).
    The electron hopping is fixed to $t_e=5t_s$. }%
    \label{fig:ARPES}
\end{figure}

The intertwined signal for the spin liquid state is, however, 
quite different from the replica scenario because the spin 
correlation is a continuum both in momenta and frequency. 
The discrepancy stems from the peculiar energy distribution 
of the photoelectric effect in Eq.~\eqref{eq:convolution}. 
The spectral function for conduction electrons can be approximated by $A_{e}(\bm{q},\omega^{\prime})\approx\delta\left(\omega^{\prime}-\xi_{\bm{q}}\right)$, in which $ \xi_{\bm{q}} $ is the difference between the electron energy $ \varepsilon_{\bm{q}} $ and the chemical potential $ \mu_{e}$. 
Equation~\eqref{eq:convolution} can be reduced to 
\begin{equation}\label{eq:reduced_convolution}
    A\left(\bm{k}, \omega \le 0\right) \propto \int\limits_{\omega\le\xi_{\bm{q}}\le0}\frac{d^{2}\bm{q}}{\left(2\pi\right)^{2}}K\left(\bm{q}\right)\mathcal{S}\left(\bm{k}-\bm{q},\omega-\xi_{\bm{q}}\right).
\end{equation}
For the integration domain, we have used the fact that both $ A_{e}\left(\bm{q},\omega\right) $ and $ \mathcal{S}\left(\bm{q},\omega\right) $ are restricted in the nonpositive frequency range $ {\omega \le 0 }$. 
Experimentally, the energy $ \xi_{\bm{q}} $ can be shifted by tuning the electron chemical potential $ \mu_e $ via gating. 
In the dilute limit, only a small number of electrons lie in the Fermi pocket around 
${ \bm{q}=\widetilde{\Gamma} }$. 
Within this pocket, $ \xi_{\bm{q}} $ can be approximated by 
$ -\delta\mu+\bm{q}^{2}/2m^{*} $, where 
$ {\delta\mu=\mu_{e}-\varepsilon_{\text{min}}} $ is the difference 
between the chemical potential $ \mu_e $ and the band minimum $ \varepsilon_{\text{min}} $, and $ m^{*} $ is the effective mass of electron. 
Moreover, the intensity modulation from the momentum-dependent 
Kondo-Hund coupling $K(\bm{q})$ can be treated as a constant $K(\widetilde{\Gamma})$ in this approximation. 
For the case $-\delta\mu>\omega$, the domain of integration 
in Eq.~\eqref{eq:reduced_convolution} can be replaced by 
${- \delta \mu \le \xi_{\bm{q}} \le 0}$ and the leading order 
term turns out to be
\begin{equation}
\label{eq: dilute limit}
	A \left(\bm{k},\omega \le 0 \right) \propto \mathcal{S} \left(\bm{k},\omega+\delta\mu\right) \delta\mu.
\end{equation} 
On the other hand, for small $\omega$ such that ${-\delta\mu<\omega}$, 
we have ${A \left(\bm{k},\omega \le 0 \right) \propto \mathcal{S} \left(\bm{k},\omega+\delta\mu\right) \delta \omega = 0}$ 
because now the frequency argument of $ \mathcal{S} $ is positive.
In any case, the information from the dynamic spin structure of the surface spin liquid can be conveyed into the intertwined spectral function as Eq.~\eqref{eq: dilute limit}.

\begin{figure}
    \centering
    \includegraphics[width=8.6cm]{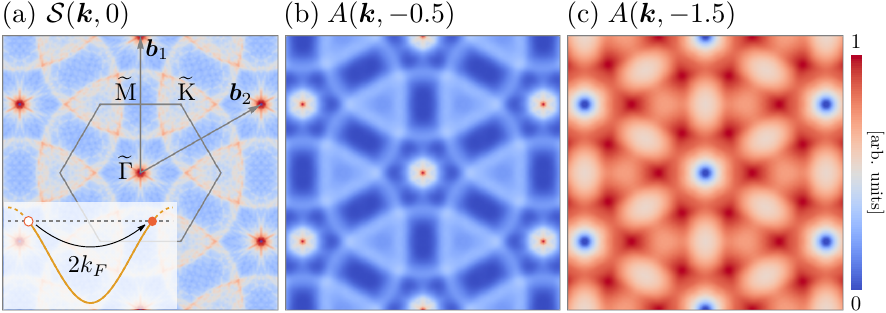}
    \caption{(a) The momentum dependence of the spin correlation $\mathcal{S}(\bm{k},0)$. The inset shows the spinon excitation with zero energy transfer, which is the origin of the enhanced intensities near the circles with radius $ 2k_F $. 
    Constant-energy images of the intertwined electron spectral function $A(\bm{k},\omega)$ are calculated with (b) $(\delta \mu, \omega) = (0.5,-0.5)$ and (c) $(\delta \mu, \omega) = (1.5, -1.5)$.}%
    \label{fig:ARPES-CE}
\end{figure}

To quantitatively demonstrate the intertwined resonance, we calculate the spectral weights for a metallic layer on top of the surface spin liquid based on the complete convolution Eq.~\eqref{eq:convolution} directly. 
For numerical convenience, the metallic layer is set to match the triangular spin superlattice of $1T$-\ce{TaS2} and they are $AA$ stacked so that the final results are uniformly modulated by the Kondo-Hund coupling $K(\bm{q})$. 
As argued previously, such a uniform modulation is always expected in the dilute limit. 
More complicated coupling modulations corresponding to distinct stackings can be simulated and compared to the experimental results similarly. 
Given that the spinon band width is much smaller than that of the conduction electrons, we take $t_s$ as the energy unit and set $t_e=5t_s$ hereafter. 
The numerical results for different electron fillings are presented in Figs.~\ref{fig:ARPES} and~\ref{fig:ARPES-CE} for high-symmetry momenta and a fixed frequency, respectively. The evolution of $ A \left(\bm{k},\omega\right) $ for more general fillings can be found in the SM~\cite{SM}.

In Fig.~\ref{fig:ARPES}(a), we present the spectral function $A_e(\bm{k},\varepsilon)$ for noninteracting electrons in the conducting layer. 
Two Fermi levels corresponding to dilute fillings with $\delta \mu = 0.5$ and $1.5$ are indicated by colored lines. 
Figure~\ref{fig:ARPES}(b) shows the spin correlation $\mathcal{S}(\bm{k},\omega)$ of the U(1) spin liquid described by the free-spinon Hamiltonian Eq.~\eqref{eq:spinon}. 
Because of the presence of the spinon Fermi surface, the particle-hole excitation can occur until the limit with zero energy transfer. 
At finite frequencies, the allowed momentum transfer excitations encode the spinon band information into the intensity of the spin correlation. 
This rich structure can be well captured by the intertwined electron spectral function $A(\bm{k},\omega)$ at the dilute limit as shown in Fig.~\ref{fig:ARPES}(c) where $\delta \mu = 0.5$. 
The slight broadening of fine structures comparing to Fig.~\ref{fig:ARPES}(b) can be attributed to the scattering between spinons and the electrons with finite energy and momentum. 
Away from the dilute limit, these scatterings become more pronounced and the detailed features of the spin correlation are inevitably blurred as shown in Fig.~\ref{fig:ARPES}(d). 
There is an overall shift of the intertwined spectral function in frequency 
when tuning the electron filling. 
The displacement is at the order of $\delta \mu$ as predicted 
by the theoretical analysis.

Near the zero frequency, the spinon Fermi surface further supports particle-hole excitations with the $2k_F$ momentum transfer as sketched in the inset of Fig.~\ref{fig:ARPES-CE}(a). 
As a consequence of such type of excitations, the momentum distribution of $\mathcal{S}(\bm{k},0)$ features pronounced overlapping circles with radius $ 2k_F $ [see Fig.~\ref{fig:ARPES-CE}(a)].
The $2k_F$ feature has also been predicted in the static spin structure factor and treated as the necessary condition for detecting spin liquids with the spinon Fermi surface~\cite{PhysRevLett.121.046401}. 
Our simulations reveal that this feature can be clearly reconstructed in the intertwined spectral function $A(\bm{k},\omega)$ after the convolution in the dilute limit, as shown in Fig.~\ref{fig:ARPES-CE}(b) where ${\delta \mu = 0.5}$. 
Considering the overall shift [compare Figs.~\ref{fig:ARPES}(b) and~\ref{fig:ARPES}(a)], we have taken the frequency ${\omega = -\delta\mu = - 0.5}$. 
Again, with the increase of the electron filling, the $2k_F$ circles become blurry, but are still discernible, as shown in Fig.~\ref{fig:ARPES-CE}(c) for $\omega = -\delta\mu = - 1.5$.

The above results suggest that the angle-resolved photoemission spectroscopy (ARPES) could measure the spin correlation function in principle. 
In real measurements, the spectrum of a pure conduction layer and a pure surface state, which can be measured separately, should be treated as the background signal of the combined system and subtracted. 
What remains is the convolution of the itinerant electron's spectral function and the spin correlation function. 
These signals may have been detected but hide in the ARPES data of $1T$-\ce{TaSe2} and $1T$-\ce{TaS2}~\cite{Chen2020,Wang2020}. 
Furthermore, raw ARPES spectra are a convolution between the electron spectral function and the momentum and energy resolution of the experiment. 
Deconvolution procedures are usually used to mitigate noises from the instruments~\cite{RAMEAU201035,doi:10.1063/1.4993919}. 
Interestingly, the itinerant electron's spectral function in Eq.~\eqref{eq:convolution} serves as a known resolution function instead. 
Similar deconvolution methods might be applied here to decode the spin correlation even with general fillings, once the intrinsic and intertwined electron spectral functions are obtained properly.

% feature of 2kF

\emph{Discussion.}---We discuss other experimental consequences of the surface spin liquid. 
The surface gapless modes give an extra contribution to the low-temperature heat capacity in this bulk-gapped material. 
It means that the powderlike samples would have a higher heat capacity than the single-crystal samples. 
As the large Kondo resonance has already been observed in Pb-doped type-II terminated surface~\cite{Shen2022}, this suggests the local moment formation and Mott physics, and further local probes such as nuclear magnetic resonance, should be helpful to detect the spin fluctuation on this surface. 
Thermal conductivity measurement can be a useful diagnosis of the gapless charge-neutral surface mode. 
There will be a magnetic contribution to thermal conductivity from type-II surface, while there is no magnetic thermal conductivity in type-I surface. 
The comparison between these two surface thermal transports can be quite insightful. 
The caveat is that the pure magnetic contribution may be difficult to obtain and may mix with the phonon transport. 
Since the monolayer $1T$-\ce{TaS2} is believed to be proximate to the Mott transition~\cite{PhysRevLett.121.046401}, it is then reasonable to expect the type-II surface to be in the weak Mott regime and the thermal Hall transport result to apply~\cite{PhysRevLett.104.066403}. 
Moreover, by varying the substrates to tune the surface electron correlations, 
it is likely to access the surface Mott transition and explore the universal 
transport properties associated with the (continuous) transition 
between the metal and surface spin liquid~\cite{PhysRevB.78.045109,SM} 
directly.

The boundary terminated \textit{exotic} liquid states are quite rare in 
condensed matter systems. 
Most were conventional states such as the Dirac-cone surface state for 
topological insulator and Luttinger liquids for quantum Hall effects~\cite{RevModPhys.82.3045,Wen1995,PhysRevB.25.2185}. 
For magnetic systems, the 2D spin-$1$ compound FeSe 
was suggested to form a coupled Haldane chain whose edge 
state would be a gapless spin-$1/2$ Heisenberg chain~\cite{Wang2015} 
and is well understood. 
A possible exotic state may emerge on the 2D surface of the spin-$1$ 
pyrochlore antiferromagnet, \ce{Tl2Ru2O7}, that was previously suggested 
as a possible formation of Haldane chains in 3D~\cite{Lee2006}. 
This would strongly depend on the orientation of Haldane chains, 
the surface lattice, and the interacting model for the emergent 
spin-$1/2$ degrees of freedom on the surface. 
There can be a triangular lattice or a kagome lattice on the (111) surface. 
In this regard, the type-II surface of $1T$-\ce{TaS2} might be the first anomalous realization of exotic quantum liquid phase. 
Finally, in addition to the surface termination, the fault surface inside the bulk $1T$-\ce{TaS2} could behave much like the type-II termination and could further support such an anomalous spin liquid.

To summarize, we explain the emergence of the spin-$1/2$ triangular lattice magnet on the type-II surface of the $1T$-\ce{TaS2}, and point out the possible existence of the spin liquid state.
This spin liquid is anomalously realized, and the experimental signatures are discussed. 

We thank Shichao Yan, Yuan Li, and Chenjie Wang for useful discussions, and Patrick Lee and Vic Law for previous collaboration. The first-principles calculations for this work were performed on Tianhe-2 supercomputer. 
We are thankful for the support from the National Supercomputing Center in Guangzhou (NSCC-GZ). This work is supported by the National Science Foundation of China with Grant No. 92065203, the Ministry of Science and Technology of China with Grants No. 2018YFE0103200 and No. 2021YFA1400300, the Shanghai Municipal Science and Technology Major Project with Grant No. 2019SHZDZX01, and the Research Grants Council of Hong Kong with General Research Fund Grant No. 17306520.

\bibliography{Ref.bib}

\end{document}

\title{Supplemental Material for ``Fractionalization on the surface: \\
Is type-II terminated 1T-TaS$_2$ surface an anomalously realized spin liquid?''}
\author{Chao-Kai Li}
\thanks{These authors contributed equally.}
\affiliation{Department of Physics and HKU-UCAS Joint Institute for Theoretical and 
Computational Physics at Hong Kong, The University of Hong Kong, Hong Kong, China}
\author{Xu-Ping Yao}
\thanks{These authors contributed equally.}
\affiliation{Department of Physics and HKU-UCAS Joint Institute for Theoretical and 
Computational Physics at Hong Kong, The University of Hong Kong, Hong Kong, China}
\author{Jianpeng Liu}
\affiliation{School of Physical Science and Technology, ShanghaiTech University, Shanghai 200031, China}
\affiliation{ShanghaiTech laboratory for topological physics, ShanghaiTech University, Shanghai 200031, China}
\author{Gang Chen}
\email{gangchen@hku.hk}
\affiliation{Department of Physics and HKU-UCAS Joint Institute for Theoretical and 
Computational Physics at Hong Kong, The University of Hong Kong, Hong Kong, China}

% \begin{abstract}

% \end{abstract}

\maketitle

\section{details of first-principles calculations}

\begin{figure}
	\includegraphics[width=17cm]{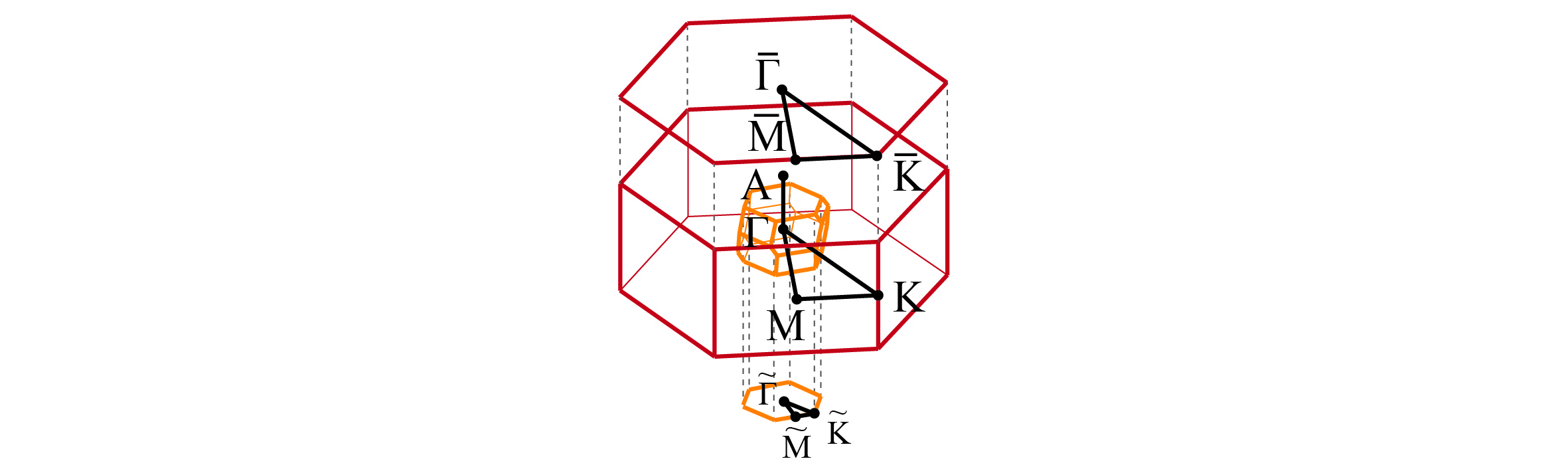}
	\caption{The Brillouin zone of the ${\sqrt{13} \times \sqrt{13} \times 2}$ supercell 
		(orange), of the primitive cell (red), and their corresponding surface Brillouin zones.}
	\label{fig: BZ}
\end{figure}

In the density functional theory calculations, the Perdew-Burke-Ernzerhof exchange-correlation functional~\cite{PhysRevLett.77.3865} and the projector augmented-wave pseudopotentials in the pslibrary~\cite{DalCorso2014,pslibrary} are used with a plane wave basis set with an energy cutoff $80$ Ry. 
In reality, the stacking sequence is partially disordered~\cite{Butler2020}, because there are symmetry-equivalent AL-type stacking fashions. 
We use the periodic structure in our calculations. 
The Brillouin zone of the $\sqrt{13} \times \sqrt{13} \times 2$ supercell is sampled by a $5 \times 5 \times 5$ grid. 
Structure optimization is performed with a convergence threshold on forces of $10^{-3} $ Ry/bohr and  a convergence threshold on pressure of $0.5$ kbar.
The Brillouin zones of the primitive cell, the supercell, and the surface are shown in Fig.~\ref{fig: BZ}. Also shown are the high symmetry points and paths.

\section{Wave function distribution of the surface states}

Comparing the spectral functions of the semi-infinite systems with type-I and type-II surfaces, one can see that the system with type-II surface has an additional band which crosses the Fermi energy [c.f. Fig.~2(b,c) in the main text]. In the following we will verify that this additional band does locate on the surface and have a look at how deeply the surface states penetrate into the bulk.

\begin{figure}
	\includegraphics[width=17cm]{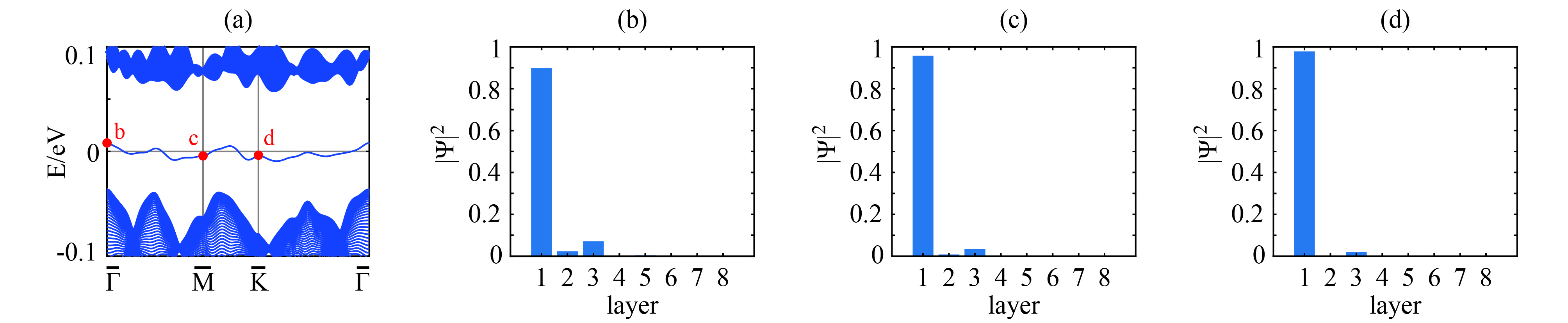}
	\caption{The band structure of the slab consisting of 100 \ce{TaS2} layers is shown in (a). The Fermi level is shifted to $ E=0 $. The electron density distributions of the states labeled with b, c, and d are shown in the subfigures (b-d), respectively. The layers are indexed from the outermost one to the bulk as 1, 2, 3...}
	\label{fig: slab_surfac_wf}
\end{figure}

After the tight-binding (TB) model of the bulk is obtained, we cut the system into a slab with 100 \ce{TaS2} layers with type-II surface on each side. The band dispersion of this slab model is shown in Fig.~\ref{fig: slab_surfac_wf}(a), which is very similar to the spectral function of the semi-infinite model in the Fig.~2(c) of the main text. The isolated band which crosses the Fermi level is clearly seen. It is two-fold degenerate due to the combination of spatial inversion and time reversal symmetry. We pick out three representative eigenstates at high symmetry points labeled by b, c, and d in Fig.~\ref{fig: slab_surfac_wf}(a), and calculate the corresponding particle density distribution over the \ce{TaS2} layers. The results are shown in Fig.~\ref{fig: slab_surfac_wf}(b,c,d), respectively. The layers are indexed from the surface to the bulk. It can be seen that the electrons are almost distributed on the outermost layer, which indicates a very small penetration depth of the surface states. It is a reflection of the large ratio of the intra- and inter-dimer hopping amplitudes. Indeed, we find $ t_{\text{intra}}^{\text{max}}/t_{\text{inter}}^{\text{max}}\approx8 $, in which $ t_{\text{intra}}^{\text{max}} $ ($ t_{\text{inter}}^{\text{max}} $) is the maximum intra-dimer (inter-dimer) hoppings.

\section{The spectral function and the spin correlation function}

In this section, we review the derivation~\cite{Sunko2020} of the Eq.~3 in the main text. We note that in Ref.~[\onlinecite{Sunko2020}], the formula is applied to the case of static magnetic order, in which only the zero frequency component of the spin correlation function is nonvanishing. In the application to the spin liquid case in this work, we need to clarify that the definition of the spin correlation function is such that it is nonvanishing for negative frequency.

The spectral function of electrons is defined by
\begin{equation}
	A\left(\vec k,\omega\right)=-\frac{1}{\pi}\text{Im}G^{\text{R}}\left(\vec k,\omega+\mi\delta\right),
\end{equation}
in which $ G^{\text{R}}\left(\vec k,\omega\right) $ is the Fourier transform of the retarded Green's function
\begin{equation}
	G^{\text{R}}\left(\vec k,t\right)=-\mi\theta\left(t\right)\sum_{\sigma}\left\langle \left\{ c_{\vec k\sigma}\left(t\right),c_{\vec k\sigma}^{\dagger}\right\} \right\rangle ,
\end{equation}
where $ c_{\vec k\sigma}^{\dagger} $ ($ c_{\vec k\sigma} $) is the electron creation (annihilation) operator, and $ \theta\left(t\right) $ is the Heaviside step function. The Green's function can be expanded by the eigenstates as
\begin{equation}
	G^{\text{R}}\left(\vec k,\omega\right)=\sum_{n\sigma}\left[\frac{\left|\left\langle \Psi_{n}^{N+1}\left|c_{\vec k\sigma}^{\dagger}\right|\Psi_{0}^{N}\right\rangle \right|^{2}}{\omega-\varepsilon_{n}^{N+1}-\mu}+\frac{\left|\left\langle \Psi_{n}^{N-1}\left|c_{\vec k\sigma}\right|\Psi_{0}^{N}\right\rangle \right|^{2}}{\omega+\varepsilon_{n}^{N-1}-\mu}\right],
\end{equation}
where $ |\Psi_{n}^{N}\rangle $ denotes the $ n $-th eigenstate of the system with $ N $ electrons, and $ \varepsilon_{n}^{N}=E_{n}^{N}-E_{0}^{N}\ge0 $ is the difference between the $ n $-th eigenenergy $ E_{n}^{N} $ and the ground state energy $ E_{0}^{N} $. For an electron system with Fermi surface, the chemical potential $ \mu=E_{0}^{N+1}-E_{0}^{N}=E_{0}^{N}-E_{0}^{N-1} $. Accordingly, the spectral function
\begin{eqnarray}
	A\left(\vec k,\omega\right)&=&\sum_{n\sigma}\bigg[\left|\left\langle \Psi_{n}^{N+1}\left|c_{\vec k\sigma}^{\dagger}\right|\Psi_{0}^{N}\right\rangle \right|^{2}\delta\left(\omega-\varepsilon_{n}^{N+1}-\mu\right)\nonumber\\&&+\left|\left\langle \Psi_{n}^{N-1}\left|c_{\vec k\sigma}\right|\Psi_{0}^{N}\right\rangle \right|^{2}\delta\left(\omega+\varepsilon_{n}^{N-1}-\mu\right)\bigg]\nonumber\\&=&A^{+}\left(\vec k,\omega\right)+A^{-}\left(\vec k,\omega\right).
\end{eqnarray}

What angle-resolved photoemission spectroscopy (ARPES) measures is the electron annihilation part of the spectral function
\begin{equation}
	A^{-}\left(\vec k,\omega\right)=A\left(\vec k,\omega\le\mu\right)=\sum_{n\sigma}\left|\left\langle \Psi_{n}^{N-1}\left|c_{\vec k\sigma}\right|\Psi_{0}^{N}\right\rangle \right|^{2}\delta\left(\omega+\varepsilon_{n}^{N-1}-\mu\right).
\end{equation}
Note that because $ \varepsilon_{n}^{N-1}\ge0 $, $ A^{-}\left(\vec k,\omega\right) $ takes nonvanishing values only for the frequency range $ \omega\le\mu $.

Now consider a special system which consists of a Mott insulating layer and a conducting layer weakly coupled to each other. Then for an energy scale much smaller than the Hubbard repulsion energy $ U $, perturbation theory can be employed to obtain an effective electron annihilation operator of the Mott insulating layer at site $ i $ with spin $ \sigma $~\cite{Sunko2020},
\begin{equation}\label{eq: effective electron operator}
	\left(c_{i\sigma}\right)_{\text{eff}}=\sum_{j\sigma^{\prime}}g_{ji}\left(\vec S_{i}\cdot\vec{\sigma}_{\sigma\sigma^{\prime}}\right)p_{j\sigma^{\prime}},
\end{equation}
in which $ \vec S_{i} $ is the spin operator of the electron at site $ i $, $ p_{j\sigma^{\prime}} $ is the electron annihilation operator at site $ j $ of the conducting layer with spin $ \sigma^{\prime} $, and $ \vec{\sigma} $ is the vector composed of Pauli matrices. $ g_{ji} $ is the coupling strength between the sites on the two layers, and is at the order of $ 1/U $. By the Fourier transformations
\begin{eqnarray}
	c_{\vec k\sigma}&=&\frac{1}{\sqrt{N}}\sum_{i}\me^{-\mi\vec k\cdot\vec R_{i}}c_{i\sigma},\\\vec S_{\vec k}&=&\frac{1}{\sqrt{N}}\sum_{i}\me^{-\mi\vec k\cdot\vec R_{i}}\vec S_{i},\\g_{\vec k}&=&\sum_{i}\me^{-\mi\vec k\cdot\left(\vec R_{j}-\vec R_{i}\right)}g_{ji},
\end{eqnarray}
we have
\begin{equation}
	\left(c_{\vec k\sigma}\right)_{\text{eff}}=\frac{1}{\sqrt{N}}\sum_{\vec q\sigma^{\prime}}g_{\vec k+\vec q}^{*}\left(\vec S_{-\vec q}\cdot\vec{\sigma}_{\sigma\sigma^{\prime}}\right)p_{\vec k+\vec q,\sigma^{\prime}},
\end{equation}
where $ N $ is the number of lattice sites.

With the effective electron creation/annihilation operators, the retarded Green's function is recast into
\begin{equation}
	G^{\text{R}}\left(\vec k,t\right)=\frac{1}{N}\sum_{\vec q\vec q^{\prime}\sigma\sigma^{\prime}\sigma^{\prime\prime}}g_{\vec k+\vec q}^{*}g_{\vec k+\vec q^{\prime}}\left[-\mi\theta\left(t\right)\right]\left\langle \left\{ \vec S_{-\vec q}\left(t\right)\cdot\vec{\sigma}_{\sigma\sigma^{\prime}}p_{\vec k+\vec q,\sigma^{\prime}}\left(t\right),\vec S_{\vec q^{\prime}}\cdot\vec{\sigma}_{\sigma^{\prime\prime}\sigma}p_{\vec k+\vec q^{\prime},\sigma^{\prime\prime}}^{\dagger}\right\} \right\rangle.
\end{equation}
As the coupling between the two layers is weak, we can treat it by a mean-field approximation,
\begin{eqnarray}
	G^{\text{R}}\left(\vec k,t\right)&=&\frac{1}{N}\sum_{\vec q\vec q^{\prime}\sigma\sigma^{\prime}\sigma^{\prime\prime}}g_{\vec k+\vec q}^{*}g_{\vec k+\vec q^{\prime}}\left[-\mi\theta\left(t\right)\right]\bigg[\left\langle \vec S_{-\vec q}\left(t\right)\cdot\vec{\sigma}_{\sigma\sigma^{\prime}}\vec S_{\vec q^{\prime}}\cdot\vec{\sigma}_{\sigma^{\prime\prime}\sigma}\right\rangle \left\langle p_{\vec k+\vec q,\sigma^{\prime}}\left(t\right)p_{\vec k+\vec q^{\prime},\sigma^{\prime\prime}}^{\dagger}\right\rangle \nonumber\\&&+\left\langle \vec S_{\vec q^{\prime}}\cdot\vec{\sigma}_{\sigma^{\prime\prime}\sigma}\vec S_{-\vec q}\left(t\right)\cdot\vec{\sigma}_{\sigma\sigma^{\prime}}\right\rangle \left\langle p_{\vec k+\vec q^{\prime},\sigma^{\prime\prime}}^{\dagger}p_{\vec k+\vec q,\sigma^{\prime}}\left(t\right)\right\rangle \bigg].
\end{eqnarray}
Because the conducting layer is nonmagnetic, the quantities such as $ \left\langle p_{\vec k\sigma}\left(t\right)p_{\vec k^{\prime}\sigma^{\prime}}^{\dagger}\right\rangle  $ should be diagonal in the spin indices, and independent on the spin channel. They should also be diagonal in the momentum indices due to the lattice translational symmetry. Furthermore, we have the identity
\begin{equation}
	\sum_{\sigma\sigma^{\prime}}\left(\vec A\cdot\vec{\sigma}_{\sigma\sigma^{\prime}}\right)\left(\vec B\cdot\vec{\sigma}_{\sigma^{\prime}\sigma}\right)=\text{tr}\left[\left(\vec A\cdot\vec{\sigma}\right)\left(\vec B\cdot\vec{\sigma}\right)\right]=\vec A\cdot\vec B.
\end{equation}
Then,
\begin{eqnarray}
	G^{\text{R}}\left(\vec k,t\right)&=&\frac{1}{N}\sum_{\vec q\sigma}K\left(\vec k+\vec q\right)\left[-\mi\theta\left(t\right)\right]\bigg[\left\langle \vec S_{-\vec q}\left(t\right)\cdot\vec S_{\vec q}\right\rangle \left\langle p_{\vec k+\vec q,\sigma}\left(t\right)p_{\vec k+\vec q,\sigma}^{\dagger}\right\rangle \nonumber\\&&+\left\langle \vec S_{\vec q}\cdot\vec S_{-\vec q}\left(t\right)\right\rangle \left\langle p_{\vec k+\vec q,\sigma}^{\dagger}p_{\vec k+\vec q,\sigma}\left(t\right)\right\rangle \bigg],
\end{eqnarray}
where we have defined $ K\left(\vec q\right)=\left|g_{\vec q}\right|^{2} $. By Fourier transformation, we get the spectral function
\begin{eqnarray}
	A_{\text{Mott}}\left(\vec k,\omega\right)&=&-\frac{1}{\pi}\text{Im}G^{\text{R}}\left(\vec k,\omega+\mi\delta\right)\nonumber\\&=&\frac{1}{N}\sum_{\vec q\sigma}K\left(\vec k+\vec q\right)\iint\frac{\dif\omega^{\prime}\dif\omega^{\prime\prime}}{\left(2\pi\right)^{2}}\bigg[\left\langle \vec S_{-\vec q}\left(\omega^{\prime}\right)\cdot\vec S_{\vec q}\right\rangle \left\langle p_{\vec k+\vec q,\sigma}\left(\omega^{\prime\prime}\right)p_{\vec k+\vec q,\sigma}^{\dagger}\right\rangle \nonumber\\&&+\left\langle \vec S_{\vec q}\cdot\vec S_{-\vec q}\left(\omega^{\prime}\right)\right\rangle \left\langle p_{\vec k+\vec q,\sigma}^{\dagger}p_{\vec k+\vec q,\sigma}\left(\omega^{\prime\prime}\right)\right\rangle \bigg]\delta\left(\omega-\omega^{\prime}-\omega^{\prime\prime}\right).
\end{eqnarray}
The subscript ``Mott'' indicates that this spectral function is calculated with the effective creation/annihilation operators~\eqref{eq: effective electron operator} of the Mott layer. The leading terms of the correlation functions in the above equation are those for the decoupled systems. Hence, it can be approximated by
\begin{eqnarray}
	A_{\text{Mott}}\left(\vec k,\omega\right)&=&\frac{1}{N}\sum_{\vec q}K\left(\vec k+\vec q\right)\int_{-\infty}^{\infty}\frac{\dif\omega^{\prime}}{2\pi}\bigg[\mathcal{S}^{+}\left(-\vec q,\omega-\omega^{\prime}\right)A_{e}^{+}\left(\vec k+\vec q,\omega^{\prime}\right)\nonumber\\&&+\mathcal{S}^{-}\left(-\vec q,\omega-\omega^{\prime}\right)A_{e}^{-}\left(\vec k+\vec q,\omega^{\prime}\right)\bigg]\nonumber\\&=&A_{\text{Mott}}^{+}\left(\vec k,\omega\right)+A_{\text{Mott}}^{-}\left(\vec k,\omega\right),
\end{eqnarray}
in which
\begin{eqnarray}
	A_{e}^{+}\left(\vec k,\omega\right)&=&\frac{1}{2\pi}\left\langle p_{\vec k\sigma}\left(\omega\right)p_{\vec k\sigma}^{\dagger}\right\rangle =\sum_{n\sigma}\left|\left\langle \Psi_{n}^{N+1}\left|p_{\vec k\sigma}^{\dagger}\right|\Psi_{0}^{N}\right\rangle \right|^{2}\delta\left(\omega-\varepsilon_{n}^{N+1}-\mu\right),\\A_{e}^{-}\left(\vec k,\omega\right)&=&\frac{1}{2\pi}\left\langle p_{\vec k\sigma}^{\dagger}p_{\vec k\sigma}\left(\omega\right)\right\rangle =\sum_{n\sigma}\left|\left\langle \Psi_{n}^{N-1}\left|p_{\vec k\sigma}\right|\Psi_{0}^{N}\right\rangle \right|^{2}\delta\left(\omega+\varepsilon_{n}^{N-1}-\mu\right)
\end{eqnarray}
are the electron spectral function of a freestanding conducting layer, with the notations explained before. And
\begin{eqnarray}
	\mathcal{S}^{+}\left(\vec q,\omega\right)&=&\left\langle \vec S_{\vec q}\left(\omega\right)\cdot\vec S_{-\vec q}\right\rangle =2\pi\sum_{n}\left|\left\langle n\left|\vec S_{-\vec q}\right|\Omega\right\rangle \right|^{2}\delta\left(\omega-\varepsilon_{n}\right)\\\mathcal{S}^{-}\left(\vec q,\omega\right)&=&\left\langle \vec S_{-\vec q}\cdot\vec S_{\vec q}\left(\omega\right)\right\rangle =2\pi\sum_{n}\left|\left\langle n\left|\vec S_{\vec q}\right|\Omega\right\rangle \right|^{2}\delta\left(\omega+\varepsilon_{n}\right)
\end{eqnarray}
are the dynamical spin correlation functions of a freestanding Mott layer, where $ |n\rangle $ is the $ n $-th eigenstate of the system, $ |\Omega\rangle = |0\rangle $ is the ground state, and $ \varepsilon_n\ge 0 $ is the difference between the $ n $-th eigenenergy and the ground state ($ n=0 $) energy. If we adopt the convention that the energy of the conducting layer is shifted so that its chemical potential $ \mu=0 $, then from the Dirac delta functions in these expressions it can be inferred that the quantities with superscript ``$ + $'' (``$ - $'') vanish for negative (positive) frequency.

As mentioned earlier, the ARPES experiments measure the the part of the spectral function below the chemical potential
\begin{eqnarray}
	A_{\text{Mott}}\left(\vec k,\omega\le0\right)&=&A_{\text{Mott}}^{-}\left(\vec k,\omega\right)\\&=&\frac{1}{N}\sum_{\vec q}K\left(\vec k+\vec q\right)\int_{-\infty}^{0}\frac{\dif\omega^{\prime}}{2\pi}A_{e}^{-}\left(\vec k+\vec q,\omega^{\prime}\right)\mathcal{S}^{-}\left(-\vec q,\omega-\omega^{\prime}\right)\nonumber\\&=&\int\frac{\dif^{2}\vec q}{\left(2\pi\right)^{2}}K\left(\vec q\right)\int_{-\infty}^{0}\frac{\dif\omega^{\prime}}{2\pi}A_{e}^{-}\left(\vec q,\omega^{\prime}\right)\mathcal{S}^{-}\left(\vec k-\vec q,\omega-\omega^{\prime}\right),
\end{eqnarray}
which is the Eq.~(3) of the main text. Note that in specifying the upper limit of the integration, we have used the fact that $ A_{e}^{-} $ vanishes for positive frequencies.

\section{The U(1) quantum spin liquid with spinon Fermi surface}

\begin{figure}
    \centering
    \includegraphics[]{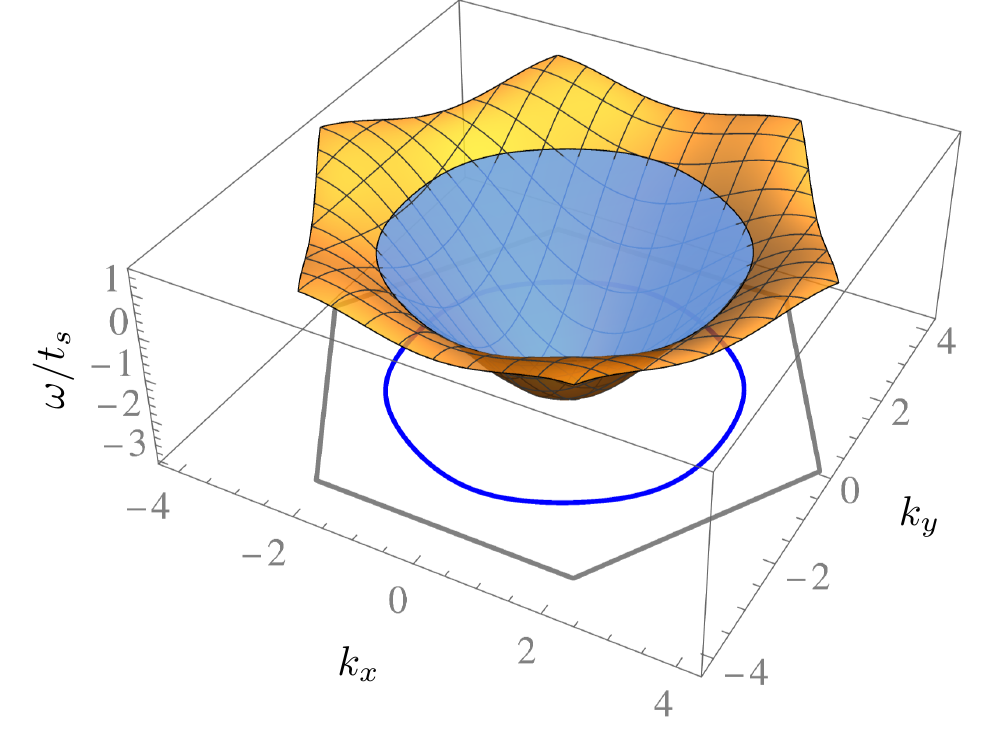}%
    \caption{The spinon mean-field band structure (orange) and the Fermi level corresponding to half filling (blue). The gray and blue lines on the $k_x$-$k_y$ plane indicate the (reduced) first Brillouin zone and the projection of the spinon Fermi surface, respectively.}%
    \label{fig:supplementary:spinonband}
\end{figure}

To demonstrate the anomalously realized spin liquid physics emerges on the type-II terminated 1T-\ce{TaS2}, we propose a mean field Hamiltonian for the spinon degrees of freedom as 
\begin{equation}\label{eq:supplementary:MFT}
    H_s=-\sum_{\braket{ij}\sigma}t_{s,ij}(f_{i\sigma}^{\dagger}f_{j\sigma}+\text{h.c.})-\mu_{s}\sum_{i\sigma}f_{i\sigma}^{\dagger}f_{i\sigma},
\end{equation}
where $t_{s,ij}$ refers to the spinon hopping on the triangular lattice whose lattice sites are located at the centers of stars of David. 
The physical spin has been represented as $\mathbf{S}_i = f_{i\sigma}^{\dagger}\frac{\bm{\sigma}_{\sigma\sigma'}}{2}f_{i\sigma'}$ where $f_{i\sigma}^{\dagger}$ and $f_{i\sigma}$ are spinon creation and annihilation operators, respectively. 
The chemical potential $\mu_s$ enforces the half filling constraint and reduces the enlarged Hilbert space to a physical one. 
We have neglected the emergent gauge field and treat the charge sector as a Mott insulator. 
Depending on the effective spin model, the real spinon Hamiltonian would be more complicated. 
Nevertheless, The mean field approximation in Eq.~\eqref{eq:supplementary:MFT} can capture the essential physics of the spinon Fermi surface. 
Without loss of generality, we only consider a uniform hopping $t_{s,ij}=t_s/2$ between nearest-neighbor sites. 
The further hopping processes would not qualitatively affect the following results. 
This leads to a two-fold degenerate spinon dispersion 
\begin{equation}
    \omega(\bm{k}) = -t_s[\cos \bm{k}\cdot \bm{a}_1 + \cos \bm{k}\cdot\bm{a}_2 + \cos \bm{k}\cdot(\bm{a}_1+\bm{a}_2) ]-\mu_s,
\end{equation}
where
\begin{equation}
    \bm{a}_1 = a(1,0), \quad \bm{a}_2 = a\bigg(-\frac{1}{2}, \frac{\sqrt{3}}{2}\bigg),
\end{equation}
are two lattice vectors for the triangular superlattice. 
We take the superlattice constant $a=1$ hereafter unless otherwise specified. 
To satisfy the half-filling constraint, the chemical potential is set to $\mu_s/t_s \approx 0.4173$. 
Moreover, the wave vectors for the Fermi level are estimated to be $k_F=2.7233$ and $k_F=2.6656$ along high-symmetry paths $\widetilde{\Gamma}$-$\widetilde{\mathrm{M}}$ and $\widetilde{\Gamma}$-$\widetilde{\mathrm{K}}$, respectively. 
In Fig.~\ref{fig:supplementary:spinonband}, we plot the spinon band structure and the corresponding Fermi surface within the reduced first Brillouin zone. 

Based on the above model, one can calculate the spin correlation function
\begin{equation}\label{eq:supplementary:correlation}
    \mathcal{S}(\bm{q},\omega) = \int_{-\infty}^{+\infty} dt e^{\imath \omega t} \braket{\mathbf{S}_{-\bm{q}}\cdot\mathbf{S}_{\bm{q}}(t)}.
\end{equation}
This quantity characterizes the response of quantum spin liquids to external stimuli such as neutrons and electrons and is believed to reveal the information of spinon fermi surface. 
For example, the $2k_F$ peaks in the static spin correlation $\mathcal{S}(\bm{q}) = \int_{-\infty}^{+\infty}\mathcal{S}(\bm{q},\omega)d\omega$ has been shown to be the necessary condition for detecting the spinon Fermi surface U(1) spin liquid. 
Formally, the spin correlation Eq.~\eqref{eq:supplementary:correlation} consists of three parts $S^{+-}$, $S^{-+}$ and $S^{zz}$. 
The former two represent the inter-band spinon particle-hole excitation and $S^{zz}$ represent the intra-band spinon particle-hole excitation. 
In general, their contributions are different and should be calculated separately. 
Fortunately, for the simple mean field Hamiltonian Eq.~\eqref{eq:supplementary:MFT} with the twofold degenerate spinon dispersion, they equally contribute to the weight of the final spin correlation and we can thus focus on one of them 
\begin{equation}\label{eq:supplementary:correlationexpand}
    \mathcal{S}(\bm{q},\omega \le 0) \propto \sum_{n} \delta(\omega_0 - \omega_n - \omega) \braket{\Omega|S_{-\bm{q}}^{+}|n}\braket{n|S_{\bm{q}}^{-}|\Omega} \propto \sum_{n} \frac{\eta/\pi\times|\braket{\Omega|S^{+}_{-\bm{q}}|n}|^2}{(\omega_0 - \omega_n - \omega)^2+\eta^2}. 
\end{equation}
Note that only half of the frequency axis where $\omega \le 0$ is considered.  
The summation takes over all intermediate particle-hole excited states $\ket{n}$ (or equivalently unoccupied states in Fig.~\ref{fig:supplementary:spinonband}) whose energies are denoted by $\omega_n$. 
The spinon ground state (occupied states in Fig.~\ref{fig:supplementary:spinonband}) and its energy are denoted by $\ket{\Omega}$ and $\omega_0$, respectively. 
The delta function introduced by the time-domain integration has been replaced by the Lorentzian distribution $\delta(\omega)=\frac{\eta/\pi}{\omega^2+\eta^2}$ with a broadening factor $\eta = 0.1 t_s$ in the following calculation. 
All other constant coefficients in the expression Eq.~\eqref{eq:supplementary:correlationexpand} are irrelevant and have been neglected for simplicity.  

\begin{figure}
    \centering
    \includegraphics[width=\textwidth]{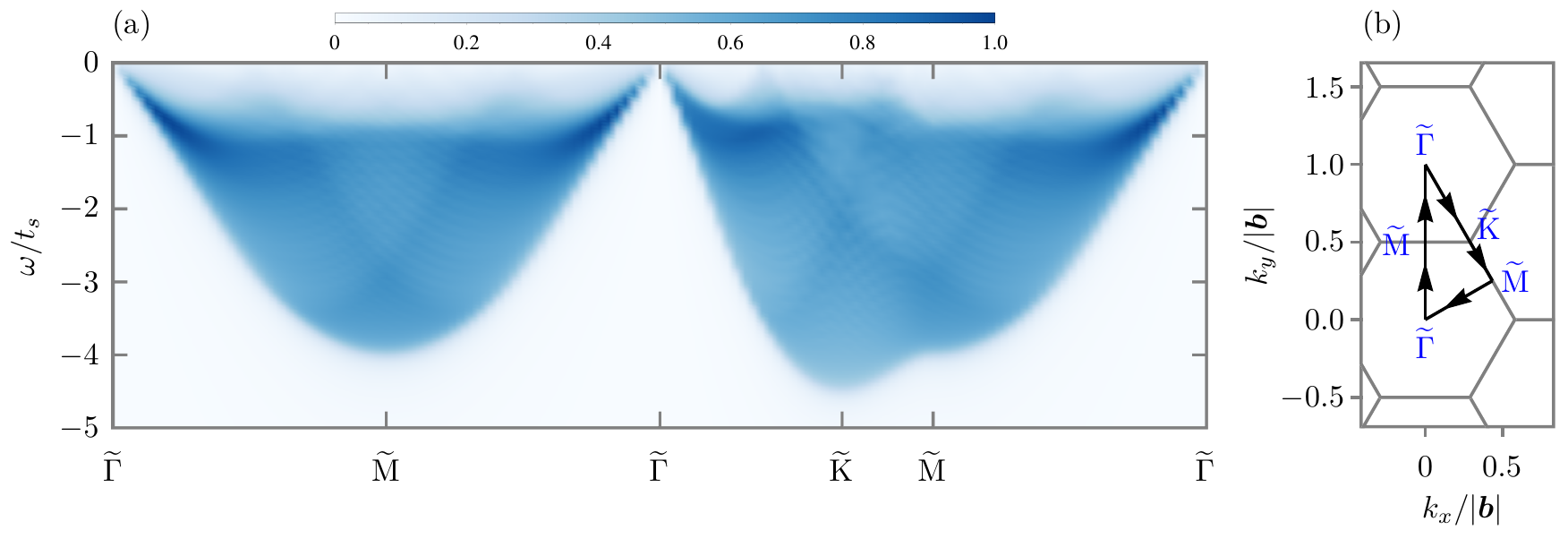}
    \caption{(a) The spin correlation $\mathcal{S}(\bm{k},\omega \le 0)$ along the high-symmetry path shown in (b) for the U(1) spin liquid with spinon Fermi surface. The portion along $\widetilde{\Gamma}$-$\widetilde{\mathrm{M}}$-$\widetilde{\Gamma}$ is identical to that shown in Fig.3(a) of the main text. Hexagons in (b) refer to the reduced Brillouin zones of a triangular lattice for stars of David on the type-II terminated 1T-\ce{TaS2}.}%
    \label{fig:supplementary:correlation}
\end{figure}

In Fig.~\ref{fig:supplementary:correlation}(a), we present the spin correlation $\mathcal{S}(\bm{q},\omega \le 0)$ along the high-symmetry path defined in Fig.~\ref{eq:supplementary:correlation}(b). 
The reduced Brillouin zone is discretized into a $64 \times 64$ mesh with respect to the reciprocal vectors $\bm{b}_{1,2}$. 
The portion along $\widetilde{\Gamma}$-$\widetilde{\mathrm{M}}$-$\widetilde{\Gamma}$ is identical to that shown in Fig.3(a) of the main text. 
At the mean field level, the spectral weight of $\mathcal{S}(\bm{q},\omega)$ would be proportional to the number of the spinon particle-hole excitations with momentum $\bm{k}$ and frequency $\omega$. 
Because the U(1) spin liquid is gapless, the spectral weights are nonzero at $\omega = 0$ except the $\widetilde{\Gamma}$ point. 
Further more, the existence of the nearly circular Fermi surface allows spinon particle-hole excitations with $2k_F$ momentum. 
These excitations have the pronounced density of states at the zero frequency and thus induce enhanced spectral weights $\mathcal{S}(2k_F,0)$ that can be identified in Fig.~\ref{fig:supplementary:correlation}. 
A more clear illustration of such physics has been present in Fig.4(a) of the main text.

\section{The intertwined electron spectral function at various fillings}

In the main text, we propose a indirect photoemission detection scheme for the anomalous realized spin liquid on the type-II terminated 1T-\ce{TaS2}. 
The proposal is base on the proximity effect between the Mott insulator surface and the conducting layer. 
The response of the itinerant electron to photoelectric effect would be modified by the underlying spin liquid in the weak Kondo/Hund's coupling limit. 
It has been proved that the leading order of the correction can be expressed as the convolution of the itinerant electron spectral function $A_e(\bm{k},\omega)$ with the spin correlation $\mathcal{S}(\bm{k},\omega)$, 
\begin{equation}
    A(\bm{k},\omega \le 0) \propto \int_{-\infty}^{0} \frac{d\omega'}{2\pi} \int \frac{d^2\bm{q}}{(2\pi)^2}K(\bm{q}) A_{e}(\bm{q},\omega')\mathcal{S}(\bm{k}-\bm{q},\omega-\omega'),
\end{equation}
where $K(\bm{q})$ is the Kondo/Hund's coupling between the itinerant electrons and the spins. 
In the main text, we have analyzed the behavior of the intertwined electron spectral function in the dilute limit. 
Over there, $A(\bm{k},\omega \le 0)$ allows a reconstruction of the spin correlation $\mathcal{S}(\bm{k},\omega)$, especially the information of the spinon Fermi surface such as the $2k_F$ signal. 
In this section, we go beyond the dilute limit and consider a general electron filling for the conducting layer. 
In the dilute limit where only a small number of electrons lie in the Fermi pocket, one can treat the Kondo/Hund's coupling $K(\bm{q})$ as a constant. 
This is not necessarily the case for general fillings; 
the momentum dependent modulation would render the weight distribution of $A(\bm{k},\omega \le 0)$ more complicated, although the reconstruction of spin correlation is still feasible in principle. 
In order to keep the demonstration simple, we assume the conducting layer and the triangular superlattice of the Mott layer are AA stacked, so that the modulation from $K(\bm{q})$ can always be uniform.

\begin{figure}
    \centering
    \includegraphics[width=\textwidth]{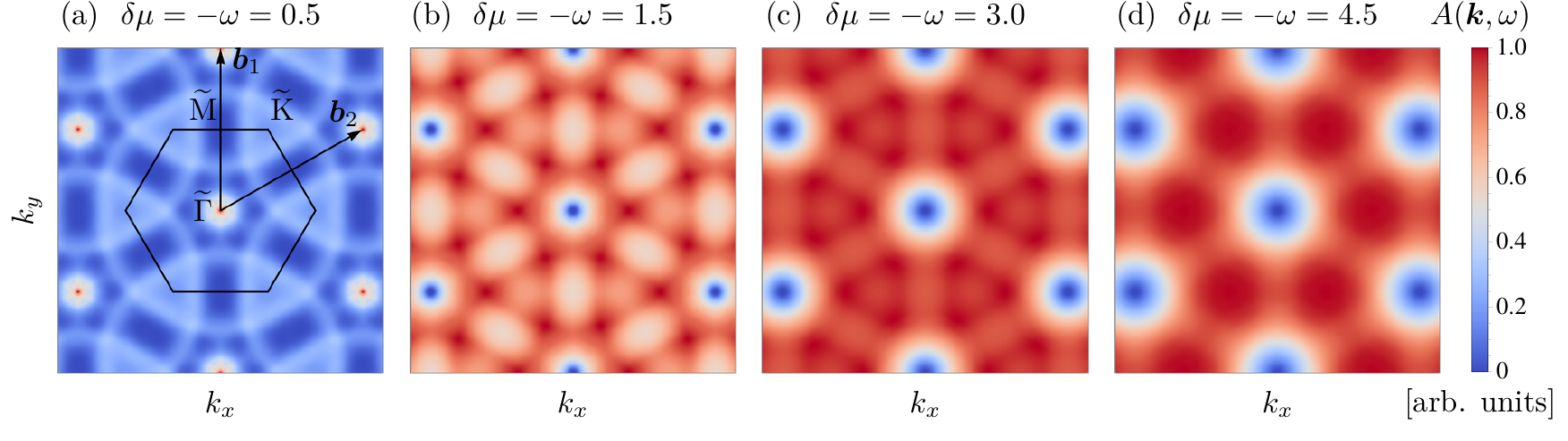}
    \caption{Constant-energy images of the intertwined electron spectral function $A(\bm{k},\omega)$ at several low fillings where $\delta \mu = - \omega$. The density is normalized in each energy level. The hexagon in (a) refers the first Brillouin zone of the triangular lattice which is formed by the stars of David. The reciprocal lattice vectors and high-symmetry momenta are indicated accordingly. The pronounced $2k_F$ features can be (a) clearly identified in the dilute limit and (b) be discernible at the very low filling. (c-d) Away from the dilute limit, these features become blurry and eventually disappear.}%
    \label{fig:supplementary:ARPES_CE}
\end{figure}

The itinerant electrons in the pure conducting layer can be described by a pure tight-binding model with the Kondo/Hund's coupling
\begin{equation}\label{eq:supplementary:electron}
    H_e = - \frac{t_e}{2}\sum_{\langle{ij} \rangle \sigma} (c_{i\sigma}^\dagger c_{j\sigma} + \text{h.c.}) - \mu_e \sum_{i \sigma}c_{i\sigma}^{\dagger}c_{i\sigma} 
    + \sum_{\langle{ik}\rangle \sigma \sigma'} K_{ik} (c_{i\sigma}^{\dagger}  \bm{\tau}_{\sigma \sigma'} c_{i\sigma'} ) \cdot \mathbf{S}_k,
\end{equation}
where $c_{i\sigma}^\dagger$ and $c_{j\sigma}$ are electron creation and annihilation operators. 
The parameters $t_e$ and $\mu_e$ are electron hopping and chemical potential, respectively. 
Without the interaction $K_{ik}$, the electron dispersion reads
\begin{equation}
    \varepsilon(\bm{k}) = -t_e[\cos \bm{k}\cdot \bm{a}_1 + \cos \bm{k}\cdot\bm{a}_2 + \cos \bm{k}\cdot(\bm{a}_1+\bm{a}_2) ]-\mu_e.
\end{equation}
The band minimum occurs at $\widetilde{\Gamma}$ and equals to $\varepsilon_{\text{min}} = - 3t_e$ for $\mu_e=0$. 
Tuning the chemical potential $\mu_e$ via gating, one can shift the maximal energy difference among occupied electrons $\delta \mu = \mu_e - \varepsilon_{\text{min}}$. 
For the half filling where $\mu_e/t_e \approx 0.4173$, the difference take $\delta \mu/t_e \approx 3.4173$. 
Because the energy scale of the itinerant electron sector is usually large than that of the spinon sector, we take the spinon hopping $t_s$ as the energy unit and set $t_e/t_s=5$ in the following simulations.

\begin{figure}
    \centering 
    \includegraphics[width=\textwidth]{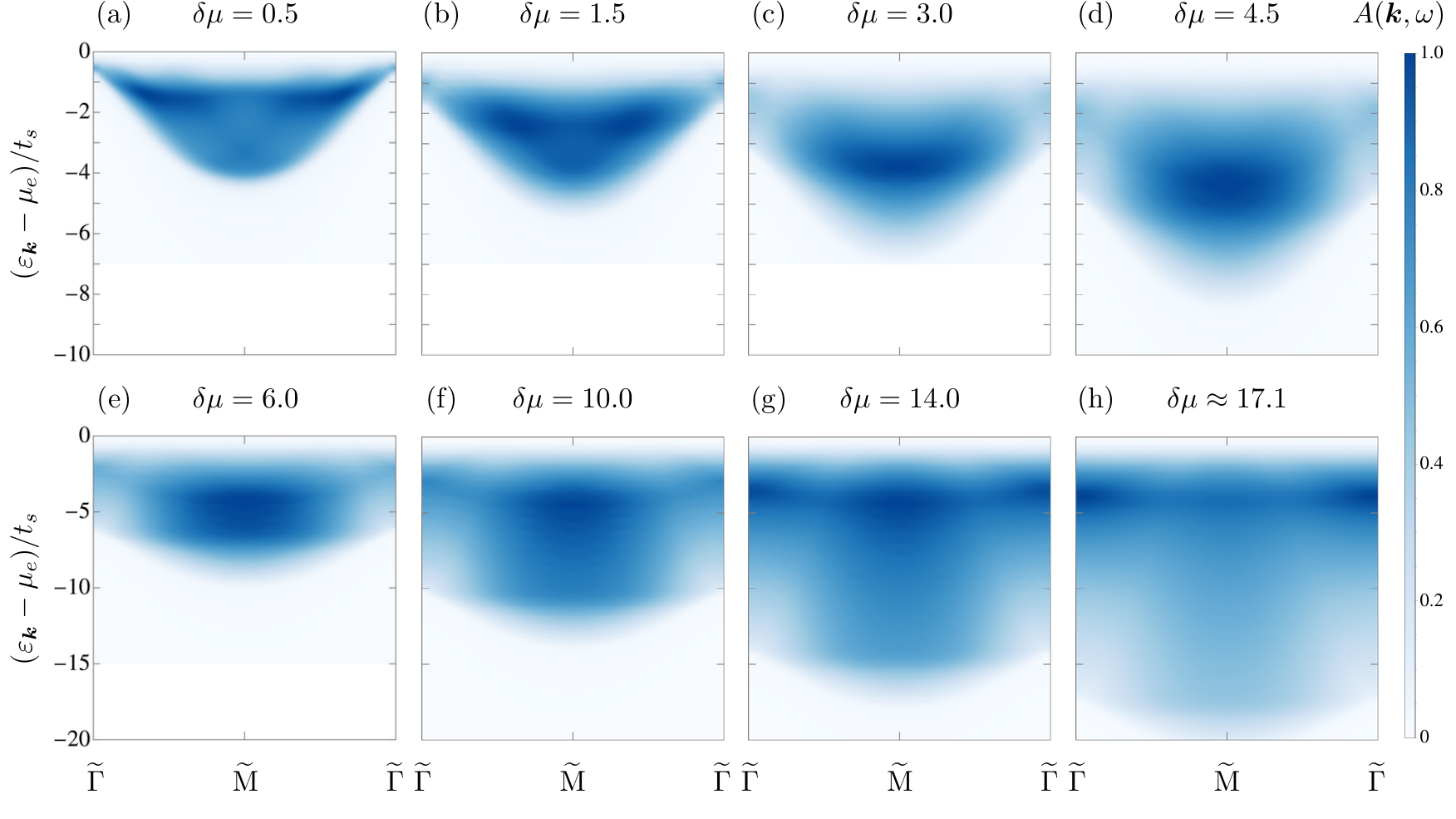} 
    \caption{The evolution of the intertwined electron spectral function $A(\bm{k},\omega)$ along the high-symmetry path $\widetilde{\Gamma}$-$\widetilde{\mathrm{M}}$-$\widetilde{\Gamma}$ from (a) the dilute limit where $\delta\mu=0.5$ to (h) the half-filling where $\delta\mu\approx 17.1$. For very low fillings, both the 
    envelope and the weight distribution of the intertwined electron spectral function $A(\bm{k},\omega)$ resemble those of the dynamic spin correlation function $\mathcal{S}(\bm{k},\omega)$. The resemblance decreases gradually with the electron filling.}%
    \label{fig:supplementary:ARPES_GM}
\end{figure}

\begin{figure}
    \centering 
    \includegraphics[width=\textwidth]{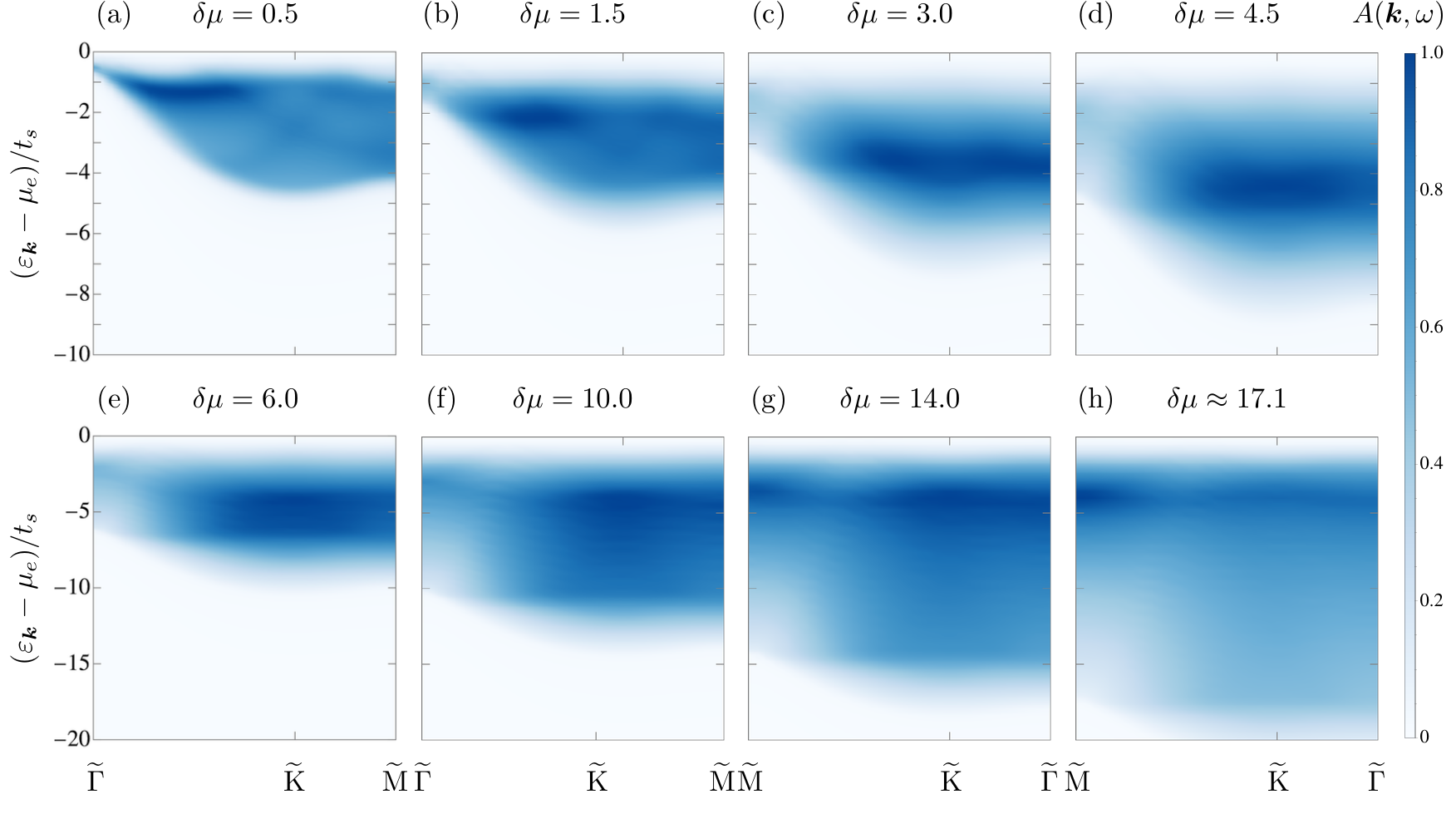} 
    \caption{The evolution of the intertwined electron spectral function $A(\bm{k},\omega)$ along the high-symmetry path $\widetilde{\Gamma}$-$\widetilde{\mathrm{K}}$-$\widetilde{\mathrm{M}}$ from (a) the dilute limit where $\delta\mu=0.5$ to (h) the half-filling where $\delta\mu\approx 17.1$. These results are similar to those in Fig.~\ref{fig:supplementary:ARPES_GM} but along different high-symmetry path.}%
    \label{fig:supplementary:ARPES_GKM}
\end{figure}

We first consider the low fillings to explore the evolution of $2k_F$ signal induced by the spinon Fermi surface. 
Because there is a small energy shift $\delta \mu$ after the convolution in the dilute limit as $A \left(\bm{k},\omega \le 0 \right) \propto \mathcal{S} \left(\bm{k},\omega+\delta\mu\right) \delta\mu$, we calculate the distributions of $A(\bm{k},\omega)$ in the reduced Brillouin zone at four different fillings and fix the energy level $\omega = - \delta \mu$. 
The results are shown in Fig.~\ref{fig:supplementary:ARPES_CE}. 
In the first two cases where $\delta \mu =0.5$ and $1.5$, the $2k_F$ signals can be clearly identified as expected. 
With the increasing of electron filling, these features are blurred and finally overwhelmed. 

In addition to the $2k_F$ signal near the low frequency, the envelope and the weight distribution of the intertwined electron spectral function $A(\bm{k},\omega \le 0)$ also resemble those of the spin correlation $\mathcal{S}(\bm{k},\omega)$ in the dilute limit, as shown in Fig.~\ref{fig:supplementary:ARPES_GM}(a) and Fig.~\ref{fig:supplementary:ARPES_GKM}(a) [c.f. Fig.~\ref{fig:supplementary:correlation}(a)].
These resemblances, however, cannot be maintained away from the low filling regime. 
In Fig.~\ref{fig:supplementary:ARPES_GM} and Fig.~\ref{fig:supplementary:ARPES_GKM}, we simulate 
the evolution of the envelope and the weight distribution of 
$A(\bm{k},\omega \le 0)$ from the dilute limit where $\delta \mu = 0.5$ 
to the half filling $\delta \mu \approx 17.1$. 
The high-symmetry paths are chosen to be $\widetilde{\Gamma}$-$\widetilde{\mathrm{M}}$-$\widetilde{\Gamma}$ and $\widetilde{\Gamma}$-$\widetilde{\mathrm{K}}$-$\widetilde{\mathrm{M}}$, respectively; 
see also Fig.~\ref{fig:supplementary:correlation}(b). 
With the increase of the electron filling, the envelope of 
$A(\bm{k},\omega \le 0)$ rapidly broadens. 
Meanwhile, the fine structures in the weight distribution, 
inherited from the spin correlation $\mathcal{S}(\bm{k},\omega)$, 
become indiscernible in the high filling regime.

\section{Discussion about surface Mott transition}

Here we give a discussion about the Mott transition of the 
surface electron from the type-II surface. 
As the bulk remains gapped throughout, the bulk electrons 
do not contribute to the low-energy physics 
at the Mott transition. Thus, one can probably
neglect the bulk electrons in the discussion about the physics
in the vicinity of the Mott transition. 
This differs from the surface singularities 
in the presence of the bulk transition from a symmetry-protected
topological state where the surface critical theory has a different
set of critical exponents from the bulk one~\cite{PhysRevLett.118.087201}.

As the bulk electron is not involved in the surface Mott transition, we 
focus the discussion on the behaviors of the surface electrons near the 
criticality. In the surface metallic phase, both the spin and charge sectors 
are gapless, while the spin sector remains gapless in the Mott regime.
This Mott transition of this effective 2D system can be described 
in the same way that was applied
 to the real 2D Mott transition between 
 a Fermi liquid metal and the spinon Fermi surface spin liquid in
 the weak Mott regime~\cite{PhysRevB.78.045109}. 
We thus expect 
zero-temperature universal resistivity jump 
as well as a
universal jump of the thermal conductivity across the Mott transition
~\cite{PhysRevB.86.245102,PhysRevB.44.6883}.

\bibliography{Ref.bib}